\documentclass[12pt]{article}

\usepackage{epsfig}
\usepackage{amssymb}
\usepackage{graphicx}
\usepackage{color}
\usepackage{cite}
\usepackage{subfigure}

\begin{document}

\baselineskip 6mm
\renewcommand{\thefootnote}{\fnsymbol{footnote}}


\newcommand{\nc}{\newcommand}
\newcommand{\rnc}{\renewcommand}



\newcommand{\tcb}{\textcolor{blue}}
\newcommand{\tcr}{\textcolor{red}}
\newcommand{\tcg}{\textcolor{green}}


\def\ba{\begin{array}}
\def\ea{\end{array}}
\def\be{\begin{eqnarray}}
\def\ee{\end{eqnarray}}
\def\nn{\nonumber\\}


\def\ct{\cite}
\def\la{\label}
\def\eq#1{(\ref{#1})}


\def\a{\alpha}
\def\b{\beta}
\def\g{\gamma}
\def\G{\Gamma}
\def\d{\delta}
\def\D{\Delta}
\def\e{\epsilon}
\def\et{\eta}
\def\ph{\phi}
\def\Ph{\Phi}
\def\ps{\psi}
\def\Ps{\Psi}
\def\k{\kappa}
\def\l{\lambda}
\def\L{\Lambda}
\def\m{\mu}
\def\n{\nu}
\def\th{\theta}
\def\Th{\Theta}
\def\r{\rho}
\def\s{\sigma}
\def\S{\Sigma}
\def\ta{\tau}
\def\o{\omega}
\def\O{\Omega}
\def\pr{\prime}


\def\half{\frac{1}{2}}

\def\goto{\rightarrow}

\def\na{\nabla}
\def\grad{\nabla}
\def\curl{\nabla\times}
\def\div{\nabla\cdot}
\def\pa{\partial}
\def\fr{\frac}

\def\bra{\left\langle}
\def\ket{\right\rangle}
\def\lb{\left[}
\def\lc{\left\{}
\def\ls{\left(}
\def\lp{\left.}
\def\rp{\right.}
\def\rb{\right]}
\def\rc{\right\}}
\def\rs{\right)}

\def\vac#1{\mid #1 \rangle}


\def\td#1{\tilde{#1}}
\def\check{ \maltese {\bf Check!}}


\def\Tr{{\rm Tr}\,}
\def\det{{\rm det}}
\def\text#1{{\rm #1}}


\def\bc#1{\nnindent {\bf $\bullet$ #1} \\ }
\def\ch {$<Check!>$ }
\def\ss {\vspace{1.5cm}}
\def\inf{\infty}

\begin{titlepage}

\hfill\parbox{5cm} { }

\vspace{25mm}

\begin{center}
{\Large \bf Holographic renormalization group flow from UV to IR}

\vskip 1. cm
   {Chanyong Park$^{a,b,c}$\footnote{e-mail : cyong21@gist.ac.kr}}

\vskip 0.5cm

{\it $^a\,$ Department of Physics and Photon Science, Gwangju Institute of Science and Technology, Gwangju 61005, Korea}\\
{\it $^b\,$ Asia Pacific Center for Theoretical Physics, Pohang 790-784, Korea } \\
{\it $^c\,$ Department of Physics, Postech, Pohang 790-784, Korea }\\

\end{center}

\thispagestyle{empty}

\vskip2cm


\centerline{\bf ABSTRACT} \vskip 4mm

We construct a three-dimensional geometry interpolating two different AdS spaces. From the dual quantum field theory viewpoint, it corresponds to a nontrivial renormalization group flow from a UV to another IR conformal field theory. On this geometry, we discuss the change of the central charge in a momentum-space and real-space renormalization group flow. The result shows in both renormalization prescriptions that the central charge monotonically decreases along the renormalization group flow.



\end{titlepage}

\renewcommand{\thefootnote}{\arabic{footnote}}
\setcounter{footnote}{0}


\section{Introduction}

After the Maldecena's AdS/CFT conjecture \cite{Maldacena:1997re,Gubser:1998bc,Witten:1998qj,Witten:1998zw}, considerable attention has been paid to holography for understanding a variety of strongly interacting systems. Since a strongly interacting system is a nonperturbative system, there is no appropriate mathematical tool to control it in the traditional quantum field theory. In this situation, the holography may be helpful to figure out the non-perturbative and quantum features of a strongly interacting system. In this short paper, we construct a dual geometry which enables us to understand holographically a non-perturbative renormalization group (RG) flow from a UV to an IR fixed point. 

Though several important features of a conformal field theory (CFT) have been well described by a one-dimensional higher AdS geometry \cite{Henningson:1998gx,Balasubramanian:1999re,deBoer:1999tgo,de Haro:2000xn,Skenderis:2002wp}, it still remains an important issue to construct a dual geometry of a deformed CFT \cite{Park:2013ana,Park:2014gja,Park:2013dqa}. In this work, we take into account a two-dimensional quantum field theory (QFT), which flows from a UV CFT to another IR CFT, and construct a dual three-dimensional gravity representing such a RG flow. Since
two fixed points at UV and IR are dual to two different AdS spaces, constructing the dual geometry of a nontrivial RG flow corresponds to find a geometry interpolating those two AdS spaces. Once such an interpolating geometry is known, we can check whether the $c$-theorem conjectured in Ref. \cite{Zamolodchikov:1986gt} is really working or not. Moreover, the dual geometry enables us to figure out how the dual QFT changes along the RG flow. After constructing a dual geometry analytically, we will investigate the change of a heavy quark's potential \cite{Maldacena:1998im,Rey:1998ik,Lee:2009bya,Park:2009nb,Park:2011zp}. The resulting quark's potential indicates that there is no phase transition when a UV CFT smoothly changes to an IR CFT. Noting that there are two different prescriptions for a RG flow. One is a momentum-space RG flow \cite{deBoer:1999tgo} and the other a real-space RG flow \cite{Ryu:2006bv,Ryu:2006ef,Mollabashi:2013lya}. In the holographic model, the momentum-space RG flow corresponds to taking the holographic renormalization procedure with regarding the radial coordinate of the dual geometry as the energy scale of the QFT. On the other hand, the real-space RG flow is governed by a holographic entanglement entropy in which the energy scale of the dual QFT is characterized by the size of the subsystem. We show that, although the $c$-function defined with two different prescriptions are different, they all satisfy the $c$-theorem. At a UV and IR fixed point, in particular, two different prescriptions reproduce the known CFT result. \\

Now, let us consider the following three-dimensional toy gravity model with $\l > 1$
\be		\la{act}
S_g= \fr{1}{2 \k^2} \int d^3 x \sqrt{-g} \ls  {\cal R} -\half \pa_\m \ph \pa^\m \ph - V(\ph) \rs,
\ee
where the scalar potential is given by
\be
V(\ph) = \frac{15 (2 \lambda -1) \cosh \left(\frac{\phi }{2}\right)-2 (2 (\lambda -1) \lambda +1) (3 \cosh (\phi )+5)+(2 \lambda -1) \cosh \left(\frac{3 \phi }{2}\right)}{16 (\lambda -1)^2 R_{\text{uv}}^2} .
\ee
This model represents a dual gravity of a two-dimensional QFT with a nontrivial RG flow from UV to  IR. Despite the  complicated form of the scalar potential, the gravity solution is given by a relatively simple and analytic solution which allows an interpolation of two different AdS geometries. \\

\noindent{\bf  UV behavior} 

Before discussing an interpolating geometric solution, we first investigate the property of the dual QFT at UV. Considering the scalar field fluctuation near the UV fixed point ($\ph=0 + \d \ph$), the scalar potential is expanded into
\be
V = - \fr{2}{R_{uv}^2} - \fr{3 \d \ph^2}{8 R_{uv}^2} + \cdots ,
\ee
which leads to a local maximum ($\pa^2 V /\pa \d \ph^2 < 0$) at $\ph=0$. Therefore, the solution of the Einstein equation at the UV fixed point $\ph=0$ reduces to an AdS space with a negative cosmological constant $\L_{uv} = V/2 = - 1/R_{uv}^2$ 
\be
ds^2 = \fr{R_{uv}^2}{z^2} \ls - dt^2 + dx^2 + dz^2 \rs ,
\ee 
where the UV region corresponds to $z \to 0$. The dual QFT of this AdS geometry becomes a two-dimensional CFT with a central charge, $c_{uv} = 12 \pi R_{uv} / \k^2$ \cite{Balasubramanian:1999re}.

If we further consider the next order correction of the scalar field fluctuation, it represents a relevant scalar deformation to the UV CFT. To see this, let us take into account a scalar field fluctuation $\d\ph$ at the UV fixed point $\ph=0$, In order to understand the asymptotic behavior of the scalar fluctuation, it is sufficient to consider the probe limit because the gravitational backreaction of the scalar fluctuation is negligible. On the UV AdS space, the scalar field fluctuation is governed by
\be
0 = \fr{1}{\sqrt{- g}} \pa_\m \ls \sqrt{-g} g^{\m\n} \pa_\n \d \ph \rs - \fr{\pa V}{\pa \d \ph}
= \frac{4 z^2 \d \phi '' -4 z \d \phi ' +3 \d \phi (z)}{4 R_{uv}^2} ,
\ee
and its solution reads
\be
\d \ph (z) = c_1 z^{1/2} + c_2 z^{3/2} .
\ee
This result shows that the scalar field fluctuation corresponds to a relevant scalar operator with the conformal dimension $3/2$. If we further consider the gravitational backreaction of the scalar fluctuation in the asymptotic region, it changes only the values of two integral constants, $c_1$ and $c_2$, but does not modify the $z$-dependence. This becomes manifest when we consider an exact solution including all gravitational backreaction later. On the dual QFT side, $c_1$ and $c_2$ are reinterpreted as a source and vacuum expectation value (vev)  of the scalar operator \cite{Gubser:1998bc,Witten:1998qj,Witten:1998zw}. \\

\noindent{\bf IR behavior} 

 For $\l > 1$, the above scalar potential also allows an additional local minimum at $\ph= \ph_{ir}$ which corresponds to an IR fixed point. If we focus only on the case with $\ph \ge 0$, the local minimum appears at
\be
\ph_{ir} = 2 \log \ls  2 \l -1 + 2  \sqrt{ \l^2 -\l} \rs ,
\ee
and near this local minima the scalar potential is expanded into
\be
V = - \fr{2 \l^2}{R_{uv}^2} +  \fr{3 \l^2 \d \ph ^2}{2 R_{uv}^2} + \cdots ,
\ee
which satisfies $\pa^2 V /\pa \d \ph^2 > 0$. In this case, the IR fixed point is again represented by another AdS space with an IR cosmological constant $\L_{ir} \equiv -1/R_{ir}^2= -\l^2/R_{uv}^2$
\be
ds^2 = \fr{R_{ir}^2}{z^2} \ls - dt^2 + dx^2 + dz^2 \rs .
\ee 
As shown in this IR solution, the parameter $\l$ in the scalar potential plays a crucial role in determining the IR fixed point. From the IR AdS geometry, we expect that the dual IR CFT has a central charge, $c_{ir}  = 12 \pi  R_{ir}/\k^2$. Comparing the central charges of UV and IR CFTs, the RG flow yields
\be
\fr{c_{ir}}{c_{uv}} = \fr{1}{\l} < 1,
\ee 
which is consistent with the Zamolodchikov's c-theorem \cite{Zamolodchikov:1986gt}.

Above, we showed that the scalar field plays a role of the relevant deformation in the UV region. Relying on the observation energy scale, its conformal dimension can vary to a different value. Near the IR fixed point, the scalar field can be represented as $\ph = \ph_{ir} + \d \ph$ and the scalar fluctuation in the probe limit is governed by
\be
0 =  \d\phi '' - \fr{1}{z} \d\phi ' -\fr{3}{z^2}   \d\phi ,
\ee
and is given by 
\be
\d \ph (z) = c_3 z^3 + \fr{c_4}{z} .
\ee
In the IR limit ($z \to \infty$), the gravitational backreaction of the first term cannot be ignored and, furthermore, can modify the background AdS geometry. As a result, $c_3$ must vanish in order to get an IR fixed point. As will be seen, a solution interpolating two UV and IR AdS geometries satisfies the constraints discussed above. \\

\noindent{\bf Renormalization group flow from UV to IR}

Let us try to find a solution interpolating two UV and IR AdS geometries. To do so, we first take the following metric ansatz
\be			\la{ans:metric}
ds^2 = \fr{R_{uv}^2 f(z)}{z^2} \ls - dt^2 + dx^2 + dz^2 \rs .
\ee
If the unkown function $f(z)$ smoothly interpolates $f(0)=1$ in the UV region and $f(\infty)= R_{ir}^2/R_{uv}^2$  in the IR region, a UV CFT continuously changes to another IR CFT by a relevant deformation. To find such a function, we look into
the equations of motion derived from \eq{act} 
\be
0 &=& \frac{f'^2}{4 f^2}-\frac{f'}{z f}+\frac{f R_{uv}^2 V(\phi)}{2 z^2}+\frac{1}{z^2}-\frac{1}{4} \phi '^2 , \nn
0 &=& \frac{f''}{2 f}-\frac{f'^2}{2 f^2}+\frac{f R_{uv}^2 V(\phi)}{2 z^2}+\frac{1}{z^2}+\frac{1}{4} \phi '^2 , \nn
0 &=& \frac{z^2 f' \phi '}{2 f^2}+\frac{z^2 \phi ''}{f}-\frac{z \phi '}{f}-R_{uv}^2 V'(\phi).
\ee
The first equation is a constraint coming from the Einstein equation and the others describe dynamics of $\ph(z)$ and $f(z)$. Since we can reconstruct the third equations by combining the other two equations, only two of them are independent.

Now, let us consider a kink-type solution which interpolates the previous UV and IR AdS spaces. We check that the following kink-type solution really satisfies two independent equations for $\l>1$ 
\be
\ph(z) &=& 2 \log \left(\frac{ (2 \lambda -1 )   z/ R_{uv}+2 \sqrt{\lambda  -1}   \sqrt{\lambda   z^2/R_{uv} ^2+ z/R_{uv} } +1}{  z/R_{uv}+1}\right) , \nn
f(z) &=&  \ls \frac{1+ z/R_{uv} }{1 + \lambda    z/R_{uv}} \rs^2 .
\ee
This analytic solution allows the asymptotic AdS geometry with the AdS radius $R_{uv}$, whereas another AdS space with the AdS radius, $R_{ir} = R_{uv}/ \l$, also appears in the IR limit ($z\to \infty$). At these two fixed points, the analytic solution reproduces the behaviors of the dual CFT mentioned before. For example, the scalar field in the UV region has the following expansion  
\be
\ph = 4 \sqrt{\lambda -1} \ls \fr{z}{R_{uv}} \rs^{1/2}-\frac{2}{3} (\lambda +2) \sqrt{\lambda -1}  \ls \fr{z}{R_{uv}} \rs^{3/2} + \cdots ,
\ee
while in the IR region it is expanded into
\be
\ph = \ph_{ir} -  \fr{4 \l (\l -1) }{2 \l - 1}  \fr{R_{ir}}{z} + \cdots ,
\ee
where the ellipsis indicates higher order corrections. These two asymptotic behaviors near the UV and IR fixed points are perfectly matched with the previous results obtained in the probe limit.

From the metric form in \eq{ans:metric}, we may reinterpret $R_{eff}=R_{uv} \sqrt{f}$ as an effective radius corresponding to the observation energy scale of the dual QFT. This is called the holographic renormalization representing the momentum-space renormalization. Then, the effective central charge of the dual QFT can be written as
\be
c_{eff} = \fr{3 R_{eff}}{2 G} ,
\ee
which reproduces the known central charges of UV and IR CFTs. At an intermediate energy scale, however, $R_{eff}$ is given by a nontrivial function of $z$, so that $c_{eff}$ crucially relies on the observation energy scale because the dual QFT is not a CFT at an intermediate energy scale. Despite this fact, $c_{eff}$ can have an important implication to the dual QFT. When a CFT at UV runs to another CFT at IR by a relevant deformation, it has been well known that there exists a $c$-function which monotonically decreases along the RG flow and that it reduces to the central charges at two fixed points. Since $c_{eff}$ reduces to the central charges of two CFTs at the fixed points, it would be interesting to check whether it monotonically decreases along the RG flow or not. To do so, let us calculate the change of $c_{eff}$ along the radial coordinate
\be
\fr{d c_{eff}}{d z} = - \fr{3  }{2 G} \fr{\l-1}{(1 + \lambda    z/R_{uv})^2} .
\ee
For $\l>1$, this is always negative. Noting that $dz$ is the RG flow direction, this result indicates that $c_{eff}$ monotonically decreases along the RG flow. Consequently, $c_{eff}$ can be taken into account a candidate of a $c$-function. Expanding $c_{eff}$ near the UV region gives rise to
\be
c_{eff} = \fr{3 R_{uv}}{2 G} -  \fr{3 (\l-1) }{2 G} \ z + \cdots  ,
\ee
whereas $c_{eff}$ in the IR regime becomes
\be
c_{eff} = \fr{3 R_{ir}}{2 G} +  \fr{3 (\l-1)  }{2 G} \fr{R_{ir}^2}{z} + \cdots  .
\ee 
Since the $c$-function describes the degrees of freedom of the dual QFT, the $c$-theorem implies that the degrees of freedom of the dual QFT monotonically decreases along the RG flow. \\

\noindent{\bf Heavy quark potential with a nontrivial RG flow}

For a CFT without a nontrivial RG flow, it has been known that the interaction of heavy quark and anti-quarks is described by the Coulomb potential inversely proportional to their distance \cite{Maldacena:1998im,Rey:1998ik}. However, since the geometry we constructed describes a nontrivial RG flow from a UV CFT to another IR CFT, it would be interesting to investigate how the heavy quark potential is modified along the nontrivial RG flow. In the string theory, a heavy quark is described by an end of an open string extended to the dual geometry. When two ends of an open string are attached to the AdS boundary, it was known that its worldsheet action represents the potential energy between two heavy quarks \cite{Maldacena:1998im}. In this case, the heavy quark's potential energy plays an important role in indicating the deconfinement phase transition of QCD. Using the holographic prescription, we will investigate the quark potential when the underlying theory changes from UV to IR.

Let us first assume that the position of the open string in the $z$-direction depends on the boundary coordinate $x$. Then, the induced metric on the open string is reduced to
\be			
ds_{in}^2 = \fr{R_{uv}^2 f}{z^2} \lb - dt^2 + \ls 1+ z'^2 \rs dx^2  \fr{}{} \rb ,
\ee
and the string action becomes
\be
S_{qq} = \fr{T R_{uv}^2}{2 \pi \a'} \int_{-l/2}^{l/2} dx \ \fr{ f \sqrt{ 1+ z'^2 } }{z^2} ,
\ee
where $T$ indicates an appropriate time interval. Using the fact that this string action has no explicit dependence on $x$, the distance between two quarks is given by
\be		\la{eq:interdistance}
d = \int_{0}^{z_0} dz \fr{2 z^2 f_0^2}{\sqrt{z_0^4 f^4 - z^4 f_0^4}} ,
\ee
where $z_0$ corresponds to the turning point and $f_0$ indicates the value of $f(z)$ at the turning point. Due to the existence of the turning point, the open string is extended only in the range of $0 \le z \le z_0$. On the other hand, the string action in terms of the turning point is given by
\be			\la{eq:qaitqaction}
S_{qq}= \fr{T z_0^2 R_{uv}^2}{\pi  \alpha' } \int_\e^{z_0} dz \ \frac{f^2 }{ z^2 \sqrt{f^4 z_0^4-f_0^4 z^4}} ,
\ee
where $\e$ is introduced to regularize the UV divergence. In the UV limit where the distance between quarks is very short, we can expand and perform the above integrals perturbatively. 

After integrating \eq{eq:interdistance}, the turning point can be perturbatively represented in terms of the quark's distance
\be
z_0 = \fr{3}{2} \ d + \fr{9 (\l-1)}{8 R_{uv}}  \ d^2 + {\cal O} \ls d^3 \rs.
\ee
The integration of the string action usually leads to a UV divergence. In order to remove such a UV divergence, we can consider an action for free quarks which also gives rise to the same UV divergence. The action of two free quarks corresponding to the free quark's mass is given by
\be
S^0_{qq} &=& \fr{T R_{uv}^2}{\pi  \alpha' } \int_\e^{\infty} dz \ \frac{f }{ z^2} \nn
&=& \frac{R_{uv}^2 T}{\pi    \alpha' \epsilon }
+ \fr{(\lambda -1) R_{uv} T}{\pi \alpha'} \ls 2  \log \epsilon  +\frac{ \left((\lambda -1)+2 \lambda  \log (\lambda/R_{uv} ) \right)}{ \lambda }  \rs,
\ee

Subtracting the free quark's mass from the string action in \eq{eq:qaitqaction}, the resulting renormalized quark potential is given by
\be
V &\equiv& \fr{S_{qq}-S^0_{qq}}{T} \nn 
&=& -\frac{2 R_{uv}^2}{3 \pi \alpha '} \ \fr{1}{d} -\frac{2 (\lambda -1) R_{uv} }{\pi  \alpha '} \ \log d +\frac{(\lambda -1) R_{uv} \left( 2 - \lambda  + 4 \l \log (2 R_{uv}/3 \lambda) \right)}{2 \pi \alpha'  \lambda } \nn
&& + \frac{ 3 (\l-1 )  (7 \l + 1)  }{8 \pi \alpha' } \ d + {\cal O} \ls d^2 \rs .
\ee
For the AdS case with $\l=1$, the quark potential reduces to a Coulomb potential, $V \sim - 1/d$, as expected by the conformal symmetry \cite{Maldacena:1998im}. For a large quark's distance, we plot the quark potential with several different $\l$ in Fig. 1. This shows that the strength of the interaction between quarks becomes weaker as $\l$ increases. Even for $\l \to \infty$, the numerical result shows that the quark's potential approaches $0$. This fact implies that there always exists a bound state of two quarks which is more stable than two free quarks. As a consequence, there is no phase transition representing the dissociation of heavy quarkonium,  when the underlying theory changes continuously from a UV CFT to another IR CFT. \\

\begin{figure}
\begin{center}
\vspace{-0.5cm}
 \includegraphics[angle=0,width=0.5\textwidth]{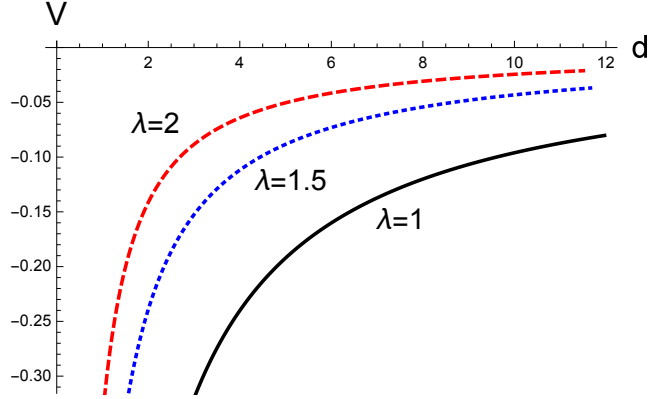}
\vspace{0cm}
\caption{\small  Quark's potential for $\l=1, \ 1.5$, and $2$.  }
\label{number}
\end{center}
\end{figure}

\noindent{\bf Entanglement entropy}

Now, let us investigate the change of the entanglement entropy along the RG flow \cite{Calabrese:2004eu,Calabrese:2005zw,Kim:2016hig,Kim:2017lyx}. The RG flow governed by the entanglement entropy is related to the real-space renormalization \cite{Park:2015hcz,Kim:2016jwu}. The entanglement entropy of the dual field is represented as the area of the minimal surface extended to the dual geometry \cite{Ryu:2006bv,Ryu:2006ef}. When we take a subsystem at $-l/2 \le x \le l/2$, the holographic entanglement entropy is described by
\be
S_E = \fr{1}{4 G} \int_{-l/2}^{l/2} dx \ \fr{R_{uv} \sqrt{ f(z)} \ \sqrt{1 + z'^2}}{z} ,
\ee
where the prime means a derivative with respect to $x$ and $G = \k^2/(8 \pi)$. Denoting the turning point of the minimal surface as $z_*$, $z'$ vanishes at this turning point and the range of $z$ extended by the minimal surface is restricted to $0 \le z \le z_*$. Solving the equations of motion derived from $S_E$, the subsystem size and the entanglement entropy can be parameterized by the turning point
\be
l &=& \int_0^{z_*} dz \ \fr{2 z \sqrt{f_*}}{\sqrt{f z_*^2 - f_* z^2}} ,  \la{eq:entanglingsize} \\
S_E &=&  \fr{R_{uv} }{2 G} \int_\e^{z_*} dz \ \frac{f  z_*}{ z \sqrt{f z_*^2-f_* z^2}} , \la{eq:entanglementen}
\ee
where $f_*$ indicates the value of $f$ at the turning point and $\e$ is introduced as a UV cutoff.

In the UV region satisfying $z_*/R_{uv} \ll 1$, performing the integral in \eq{eq:entanglingsize} perturbatively leads to
\be
l =  z_* - \fr{\l-1}{R_{uv}} z_*^2  +{\cal O} (z_*^3) .
\ee
The inverse of this relation determines the turning point in terms of the subsystem size
\be
z_* = l +  \fr{\l-1}{R_{uv}}  l^2 +{\cal O} (l^3) .
\ee
The integral in \eq{eq:entanglementen} leads to
\be
S_E =\frac{ R_{{uv}}  }{2 G} \log \frac{l}{\e}  +\frac{R_{uv}}{8 G} -\frac{ (\lambda -1) }{4 G} l +{\cal O} (l^2) .
\ee
Here the first term is exactly the result of a two-dimensional CFT and the last term represents the nontrivial contribution caused by the scalar deformation. This result shows the UV fixed point ($l \to 0$) is not stationary because $d S_E/d l \ne 0$. The central charge of the dual CFT is determined by \cite{Casini:2004bw,Myers:2010tj,Casini:2012ei}
\be
c_E  \equiv 3 \fr{d S_E}{d \log l} = \frac{ 3 R_{{uv}}  }{2 G}     -\frac{ 3 (\lambda -1) }{4 G} l  + \cdots .
\ee
$c_E$ defined here is another candidate of a $c$-function represented in the real-space RG flow. In the UV limit ($l \to 0$), as expected, $C$ reduces to the known CFT's central charge, $c_{uv} = 3 R_{uv}/ (2 G)$.

\begin{figure}
\begin{center}
\vspace{-0.cm}
\subfigure[]{\includegraphics[angle=0,width=0.45\textwidth]{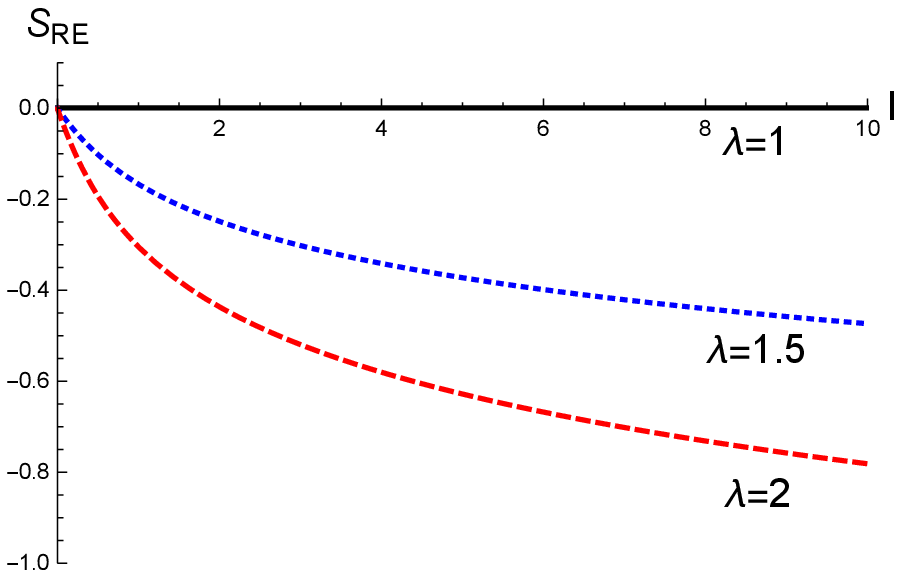}}
\hspace{0.5cm} 
\subfigure[]{\includegraphics[angle=0,width=0.45\textwidth]{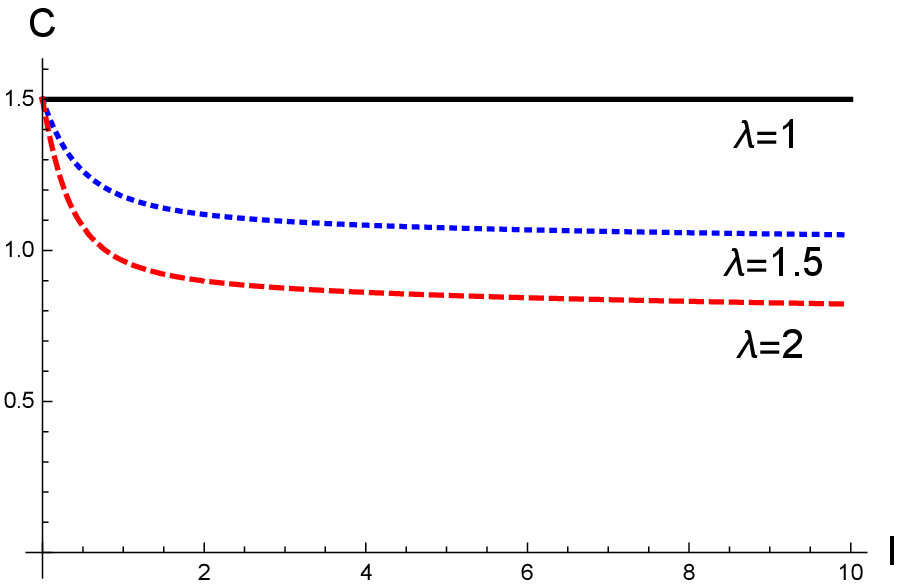}}
\vspace{0cm}
\caption{\small  The RG flow of (a) the renormalized entanglement entropy, $S_{RE} \equiv S_E - \fr{R_{uv}}{2 G} \ls \log \fr{l}{\e} + \fr{1}{4}\rs$, and (b) the $c$-function for $\l=1, \ 1.5$, and $2$ where we take $G=1$ and $R_{uv}=1$ for simplicity.  }
\label{number}
\end{center}
\end{figure}

In Fig. 2, we depict the change of the entanglement entropy and the $c$-function with several $\l$s at the intermediate energy scale. The results show that the $c$-function defined by the entanglement entropy also monotonically decreases along the RG flow, which is again consistent with the Zamolodchikov c-theorem. Moreover, the result in Fig. 2(b) shows that the $c$-function approaches to the central charge of the IR CFT, $c_{ir} = 3 R_{ir}/ (2 G)$, when the subsystem size becomes large ($l \to \infty$). At the fixed points, we have obtained the exact same central charges consistent with the CFT results regardless of the RG flows defined in momentum and real spaces. This is because of the restoration of the conformal symmetry under which the momentum scales as the inverse of the coordinate, $p \sim 1/l$. Therefore, there exists a direct relation between the momentum and coordinate at the fixed points. However, since the conformal symmetry is broken at the intermediate energy scale, there is no simple relation between $p$ and $l$. Despite this fact, the holographic RG flows in the momentum- and real-space consistently show that the $c$-function defined in a different way perfectly satisfies Zamolodchikov c-theorem.\\

In this work, we have discussed the $c$-function of a two-dimensional QFT along the nontrivial RG flow. In spite of the importance of the non-perturbative RG flow to understand the IR features, it is not easy to know the non-perturbative RG flow of a higher dimensional QFT in a traditional QFT. In this situation, the holographic method we investigated here would be useful to account for the non-perturbative IR phenomena of nuclear and condensed matter physics. We hope to report more results on higher dimensional theories in future works.

\vspace{1cm}

{\bf Acknowledgement}

C. Park was supported by Basic Science Research Program through the National Research Foundation of Korea funded by the Ministry of Education (NRF-2016R1D1A1B03932371) and also by the Korea Ministry of Education, Science and Technology, Gyeongsangbuk-Do and Pohang City.

\vspace{1cm}


\bibliography{bib}{}

\end{document}